\documentclass{aa}
\usepackage{graphicx}
\usepackage{txfonts}
\begin{document}
   \title{On shear-induced turbulence in rotating stars}


  \author{S. Mathis         
	 \inst{1}          
	 \and          
	 A. Palacios          
	 \inst{2}          
	 \and          
	  J.-P. Zahn          
	  \inst{1}
	  }

   \offprints{J.-P. Zahn}

   \institute{LUTH, Observatoire de Paris, F-92195 Meudon\\
              \email{stephane.mathis@obspm.fr; jean-paul.zahn@obspm.fr}
         \and
              Inst. Astron. Astrophys., Univ. libre Bruxelles, CP 226 blvd. du Triomphe, B-1050 Brussels\\
             \email{palacios@astro.ulb.ac.be}  \\       
            }

   \date{Received February 17; accepted March 3, 2004}

   \abstract{
We review various prescriptions which have been proposed for the turbulent transport of matter and angular momentum in differentially rotating stellar radiation zones. A new prescription is presented for the horizontal transport associated with the anisotropic shear turbulence which is  produced by the differential rotation in latitude; this `$\beta$-viscosity' is drawn from torque measurements in the classical Couette-Taylor experiment (Richard \& Zahn 1999). Its implementation in a stellar evolution code leads to enhanced mixing, as illustrated by models of a rotating main-sequence star of 1.5 solar mass. 
     \keywords{turbulence --
                stars: evolution --
                stars: rotation
               }
   }

   \maketitle
%

\section{Introduction}

In standard  models of stellar interiors, radiation zones, which are convectively stable, are postulated to be without motion other than rotation, except in the vicinity of convection zones, where often convective overshoot is accounted for in a crude way. But it is well known, since Eddington (1925) and Vogt (1925), that in rotating stars the centrifugal force breaks the radiative equilibrium, and thus causes a slow large-scale circulation, which transports matter and angular momentum. A similar circulation is induced locally in the vicinity of a differentially rotating convection zone, in the so-called tachocline. As a result, radiation zones rotate non-uniformly, and therefore they are liable to various instabilities which may generate turbulence. 

This turbulence contributes to the transport of matter and angular momentum, and to model these in a stellar evolution code, one needs reliable prescriptions for the turbulent transport. To derive such prescriptions remains today one of the most challenging problems, not only in stellar physics, but also in oceanography and in atmospheric sciences. The reason is that turbulence is still poorly understood, particularly when it departs from ideal conditions: homogeneity, isotropy, stationarity. ``Turbulence is an order of magnitude subject; there is no rigorous theory of turbulence'' warned M. Hoffert at the Aspen Global Change Institute in 1997. And he concluded:  ``The way turbulence is represented in current (global circulation) models is not realistic and this has important ramifications for the climate sensitivity predicted by these models.''

We could use the same words to describe the situation when modeling stellar interiors, replacing ``climate sensitivity'' by ``evolutionary sequences''. Therefore it takes some courage to - nevertheless - engage in this task, as we shall attempt in this article. We shall restrict our scope on turbulence produced by shear instabilities, which are likely to be the most powerful, at least in a non-magnetic star. 

Why do we believe that stellar radiation zones are turbulent? Some velocity profiles are known to be linearly unstable, as soon as the Reynolds number characterizing the flow exceeds some critical value. (We recall that this number involves the typical size $L$ and and the typical velocity contrast $V$ of the flow: $Re = V L/ \nu$, where $\nu$ is the kinematic viscosity; it measures the relative strength of advection and viscous transport.) But flows that are linearly stable are found turbulent in laboratory experiments, again above some critical Reynolds number $Re_c$, due to non-linear processes which are still not completely elucidated (see for instance Richard \& Zahn 1999).  Owing to their huge size and their low viscosity, stellar radiation zones have a Reynolds number which easily exceeds the critical values measured in the laboratory, $Re \gg Re_c$, and therefore they should be turbulent, according to this simple criterion. 

What seems more powerful in preventing a shearing flow from becoming unstable is the buoyancy force in a stable stratification, or the Coriolis force for fast enough rotation. We therefore expect these forces to play an important role in controlling the turbulence in stellar radiation zones; in particular, the turbulent motions will probably be anisotropic, with stronger transport in the horizontal than in the vertical direction.

Several papers have been devoted to the description of such shear turbulence, and to the derivation of prescriptions for the turbulent transport; to quote the most recent ones we refer to Zahn (1992),  Maeder (1995, 1997), Talon and Zahn (1997), Schatzman et al. (2000), Maeder (2003). Here we shall present an update of the subject, and recall the main results. We shall also introduce a new prescription for the horizontal transport, which is drawn from laboratory experiments, and discuss its impact on a stellar model, namely that of an evolving 1.5 M$_\odot$ star.

\section{Turbulence induced by the vertical shear}

It is well established that a stable entropy stratification acts to inhibit 
the instability caused by a vertical shear $V(z)$, 
where $V$ is the amplitude of the horizontal velocity and $z$ the
vertical coordinate. 

In the adiabatic case, i.e. in the absence of thermal leakage and viscous dissipation,
a necessary condition for shear instability is
expressed by the Richardson criterion:
\begin{equation}\label{eq:richad}
{N^2 \leq \left({{\rm d} V \over {\rm d}z}\right)^2} Ri_{\rm c}.
\end{equation}
It states that the shear rate ${\rm d}V/{\rm d}z$ has to overcome the
stabilizing effect of the buoyancy force, which  is measured by  
 the Brunt-V\"ais\"al\"a frequency $N$:
\begin{equation}\label{eq:BV}
N^2 = N^2_{T} + N^{2}_{\mu} = 
{g \delta \over H_P} \left(\nabla_{\rm ad} - \nabla \right) +
{g \varphi \over H_P} \nabla_{\mu}.
\end{equation}
We employ the usual notations for temperature $T$, density $\rho$, 
pressure $P$, molecular weight $\mu$, 
gravity $g$, pressure scale height $H_P$, and for
the logarithmic gradients 
$\nabla = {\rm d} \ln T /{\rm d} \ln P$
 and  $\nabla_{\mu} = {\rm d} \ln \mu/ {\rm d} \ln P$;
$\nabla_{\rm ad} = (\partial \ln T / \partial \ln P)_{\rm ad}$ is the adiabatic temperature gradient.
   The coefficients $\delta 
= - (\partial \ln \rho / \partial \ln T)_{P, \mu}$ and
$\varphi = (\partial \ln \rho / \partial \ln \mu)_{P, T}$ are drawn 
from the equation of state; they are unity for perfect gas.  
The critical Richardson number $Ri_{\rm c}$ is of order unity for typical 
flow profiles; in the following we shall take $Ri_{\rm c} = 1/4$.

However, when the turbulent eddies exchange heat with their 
environment, the stabilizing effect of buoyancy is reduced, 
as was shown by Townsend (1958). In an optically thick medium,
this reduction is determined by the P\'eclet
number characterizing the turbulence:
\[Pe=\frac{v \ell}K,\]
where $v$ and $\ell$ are the velocity and the size of the 
largest turbulent elements,
and $K$ the thermal diffusivity:
\[K=\frac{16}3\frac{\sigma T^3}{C_{P}\rho^{2}\kappa },\]
$\sigma$ is the Stefan constant,
$\kappa$ the Rosseland mean opacity and $C_{P}$ the heat capacity
at constant pressure.

When ${v \ell} \ll K$ thermal diffusion proceeds faster than 
advection and, if there is no composition gradient, 
the Richardson criterion for instability takes the modified form
(Zahn \cite{z74}):
\begin{equation}\label{eq:richmod}
 {v \ell \over K} N_{T}^2 < Ri_{\rm c} \left({{\rm d} V \over {\rm d}z}\right)^2 .
 \label{ri-mod}
\end{equation}
Identifying the largest $v \ell$ satisfying this inequality with the turbulent diffusivity $D_t$, one obtains
\begin{equation}\label{eq:diff-t}
 D_t =   K \, {Ri_{\rm c}\over  N_{T}^2} \left({{\rm d} V \over {\rm d}z}\right)^2 .
\end{equation}
This result is consistent with the linear stability analysis of a
hyperbolic-tangent profile imbedded in a stable stratification, including thermal diffusion, which were performed by Dudis (\cite{d74}) and more recently by Ligni\`eres et al. (1999). 

In presence of a vertical composition gradient, only the thermal component
of the buoyancy force is weakened through radiative diffusion, and the
instability criterion would read then (Zahn 1974; Maeder 1995)
\begin{equation}\label{eq:BV-mu}
{v \ell \over K} N^2_{T}  + N^{2}_{\mu} < Ri_{\rm c} \left({{\rm d} V \over {\rm d}z}\right)^2 .
\end{equation}
Meynet and Maeder (1996) pointed out that this condition was too severe, and that it would not allow for the mixing which is observed in massive stars. 

But the $\mu -$component of the buoyancy can also be reduced by the turbulence. This is true in particular when this turbulence is anisotropic, with much stronger transport in the horizontal than in the vertical direction, as expected here due to the stable stratification. Then $v$ and $\ell$ characterize the largest eddies which are still isotropic, and the horizontal component of the turbulent diffusivity is much larger than the vertical one: $D_h \gg v \ell$. It was argued by Talon and Zahn (1997) that such turbulence erodes the fluctuations of chemical composition which provide the restoring force, and that
the Richardson criterion then takes the form
\begin{equation}\label{eq:BV-final}
{v \ell \over (K + D_h)} N^2_{T}  + {v \ell \over D_h} N^{2}_{\mu} < Ri_{\rm c}\left({{\rm d} V \over {\rm d}z}\right)^2 .
\end{equation}
The rotation rate $\Omega$ does not appear in this inequality because the buoyancy force is stronger than the Coriolis force, $N \gg \Omega$. 
From (\ref{eq:BV-final}) it is straightforward  to derive the vertical component $D_v = v \ell$ of the turbulent diffusivity: 
\begin{equation}\label{eq:dv-mu}
D_v = {  Ri_{\rm c}  \over  N^2_{T} /(K + D_h) +  N^{2}_{\mu}/ D_h}  \left(r {{\rm d} \Omega \over {\rm d}r}\right)^2 ;
\end{equation}
we have replaced the shearing rate by its expression in spherical geometry: ${\rm d}V/{\rm d}z \rightarrow r \sin \theta \, {\rm d}\Omega /{\rm d}\theta$, and performed a spherical average. The critical Richardson number has been renormalized; its canonical value is now $Ri_c=1/6$. A similar expression is being used by Maeder and Meynet (2001).

\section{Turbulence induced by the horizontal shear}

Differential rotation in latitude, $\Omega(\theta)$, is known to be linearly unstable when one of the following conditions is met. Either the specific angular momentum $(r \sin \theta)^2 \Omega$ has to decrease from pole to equator, on a level surface, according to the H{\o}iland-Solberg criterion, a situation which is highly improbable in stars. Or the vorticity profile $\partial (\sin^2\theta \, \Omega) / \sin \theta \, \partial \theta$ must present a maximum in $\theta$, a criterion akin to the famous inflexion point theorem (Rayleigh 1880), which has been transposed by  Watson (1981) to spherical geometry. This requirement is probably not satisfied in stars, except perhaps in a tachocline (see Garaud 2001).

But in stars the Reynolds number characterizing such differential rotation is easily so large that it is tempting again to conclude that even a slight shear, although linearly stable,  is prone to {\it non-linear} instabilities.
However some caution must be exerted here, since the Coriolis force has a stabilizing effect on shear instabilities, and the role of curvature remains to be fully elucidated. Although work is in progress on that subject, most of what is known so far is based on laboratory experiments. There turbulence tends to suppress the cause of the instability, namely the differential rotation when the turbulence is generated by the shear (Richard \& Zahn 1999). We shall therefore assume that the same occurs in stellar radiation zones, and that the main role of the turbulent viscosity $\nu_h$ is to reduce the horizontal shear produced by the advection of angular momentum  through the meridional circulation.
This is described by the advection/diffusion equation governing the differential rotation in latitude:
\begin{equation}\label{eq:om2}
\rho {{\rm d} \over  {\rm d} t} (\rho r^2 \Omega_2) -2 \rho \overline \Omega r 
[2 V_2 - \alpha  U_2] = - 10 \rho \nu_h \Omega_2,
\end{equation}
where the meridional circulation is given by 
\begin{equation}
u_r(r, \theta)=U_2(r) P_2(\cos \theta), \quad u_\theta(r, \theta)=V_2(r){\rm d}P_2/{\rm d}\theta,
\end{equation}
\begin{equation}
\hbox{with} \; V_2(r) = {1 \over 6 \rho r} {{\rm d}  \over {\rm d} r} \left(\rho r^2 U_2\right),  \quad
\alpha(r) = {1 \over 2} {{\rm d} \ln  r^2 \overline\Omega \over {\rm d}  \ln r} ,
\end{equation}
and where we have expanded the angular velocity in
$\Omega(r, \theta) =  \overline \Omega(r) + \Omega_2(r) [P_2(\cos \theta)+1/5]$\footnote{The constant 
$1/5$ is missing in Zahn (1992) and Maeder \& Zahn (1998); the correct expression  is derived in Mathis \& Zahn (2004), as well as eq. (\ref{eq:om2}).}, 
the mean angular velocity being defined as 
\begin{equation}
\overline\Omega(r) = {\int_0^\pi \Omega(r, \theta) \sin ^3 \theta {\rm d}  \theta \over
\int_0^\pi \sin ^3 \theta {\rm d}  \theta} .
\end{equation}
In the stationary limit, advection and diffusion balance each other:
\begin{equation}\label{eq:om2stat}
 \overline \Omega r [2 V_2 - \alpha  U_2] =  5 \nu_h \Omega_2 .
\end{equation}

For lack of a better prescription, it was proposed in Zahn (1992) that $\nu_h$ be determined by the condition that it enforces a small, postulated amount of differential rotation $\Omega_2/\overline\Omega$. Then 
\begin{equation}\label{eq:nuh-92}
 \nu_h = {1 \over c_h}  r [2 V_2 - \alpha  U_2] ,
\end{equation}
where $c_h \leq 1$ is a parameter of order 1,
with a similar expression for the turbulent diffusivity $D_h$.

The horizontal diffusivity due to that shear turbulence interfers with the transport of chemicals through the meridional circulation, as was described by Chaboyer and Zahn (1992). When the turbulence is sufficiently anisotropic, the combination of both processes (meridional advection + horizontal diffusion) results in a vertical diffusion whose coefficient is given by
\begin{equation}\label{eq:deff}
D_{\rm eff} = {1 \over 30} {(rU)^2 \over D_h} ,
\end{equation}
and which competes with that, $D_v \approx \nu_v$, caused by the vertical shear.

As Maeder (2003) has shown very convincingly, the crude prescription (\ref{eq:nuh-92}) given above allows for horizontal fluctuations of the molecular weight which are too large and tend to stop the meridional circulation. It also leads to a vertical diffusivity $D_v$, according to (\ref{eq:dv-mu}), which is insufficient to account for the mixing observed in massive stars. Moreover, it does not guarantee that $D_h \gg D_v$, as is assumed in deriving the effective diffusivity $D_{\rm eff}$. 

For these reasons, Maeder proposes another prescription for this horizontal shear viscosity, which is based on the balance between the rate at which kinetic energy is channeled into differential rotation by the horizontal component of the meridional circulation, and the rate of viscous dissipation. He assumes that the dynamical time involved in the first process can be approximated by the time needed by a lagrangian particle drifting from equator to pole to be wound up by (say) $2 \pi$. This yields a horizontal viscosity 
\begin{equation}\label{eq:mae-03}
 \nu_h = A  r (r \overline \Omega V_2 [2 V_2 - \alpha  U_2])^{1/3} ,
\end{equation} 
with $A$ of order $1/10$.

Here we take another approach, and borrow a prescription for the turbulent transport from laboratory experiments.
Cylindrical Couette-Taylor flow has been found turbulent in the laboratory when the rotation rate $\Omega$ increases with the distance $s$ from the axis, whereas the linear instability analysis predicts that the flow should be stable. Experiments performed with the inner cylinder at rest, and varying the size $\Delta s$ of the gap between cylinders, have shown that when the relative gap $\Delta s / s$ is larger than about $1/20$, turbulence sets in whenever the gradient Reynolds number $s^3 ({\rm d} \Omega/{\rm d} s)/\nu$ exceeds some critical value.  A turbulent viscosity can be drawn from the torque measurements, and it is found to scale as
\begin{equation}\label{eq:nu_RZ}
\nu_t = \beta s^3 \, {{\rm d} \Omega \over {\rm d} s} ,
\end{equation}
with $\beta \approx 1.5 \, 10^{-5}$ (Richard \& Zahn 1999).
This `$\beta$-viscosity' has been applied to accretion disks by Hur\'e et al. (2001), where it yields somewhat different results compared to the classical $\alpha$-viscosity introduced by Shakura \& Sunyaev (1973).

This prescription has been established in the case of maximum differential rotation;  it remains to be verified whether its validity extends to milder shear rates, which are typical of stellar interiors. Transposing (\ref{eq:nu_RZ}) to spherical geometry, with the shear in latitude only,
\begin{equation}
\nu_t = \beta (r \sin \theta)^3 \left| {{\rm d} \Omega\over r {\rm d}  \theta} \right| = 
\beta r^2 |\Omega_2| \sin^3  \theta \left| {{\rm d} P_2 \over {\rm d}  \theta} \right|,
\end{equation}
and averaging over the sphere, we obtain a turbulent diffusivity
\begin{equation}\label{eq:nu_RZsph}
\nu_h = {1 \over 2} \beta r^2 | \Omega_2| .
\end{equation}
Inserting this value in (\ref{eq:om2stat}), we get the following expression for the turbulent viscosity
\begin{equation}
\nu_h = \left({\beta \over 10}\right)^{1/2} (r^2 \overline \Omega)^{1/2} [r |2 V_2 - \alpha  U_2|]^{1/2} ,
\label{new-nuh}
\end{equation}
again in the stationary limit. The same value will be used for the turbulent diffusivity $D_h$.

 \begin{figure}
   \centering
   \includegraphics[width=8cm]{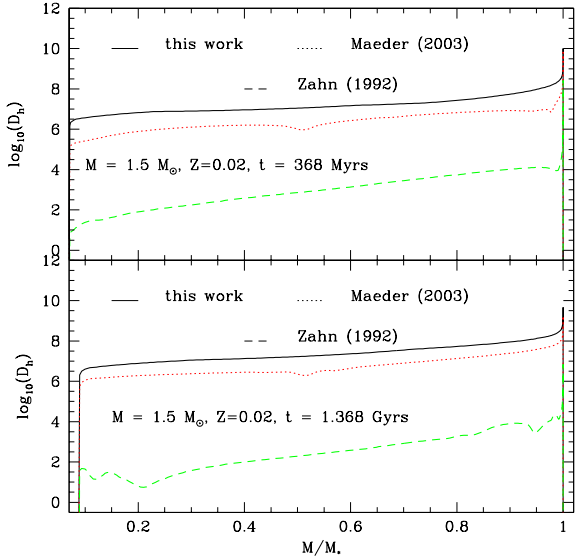}
      \caption{Horizontal component $D_h$ of  the turbulent diffusivity generated by the horizontal shear, in a 1.5 M$_\odot$ star rotating initially with an equatorial velocity of 110 km s$^{-1}$, at two evolutionary stages (0.368 and 1.368 Gyr). Comparison of the old prescription (eq. \ref{eq:nuh-92};  Zahn 1992) with that of eq. \ref{eq:mae-03} (Maeder 2003) and that of the present paper (eq. \ref{new-nuh}). 
              }
         \label{fig.1}
   \end{figure}
   
    \begin{figure}
   \centering
   \includegraphics[width=8cm]{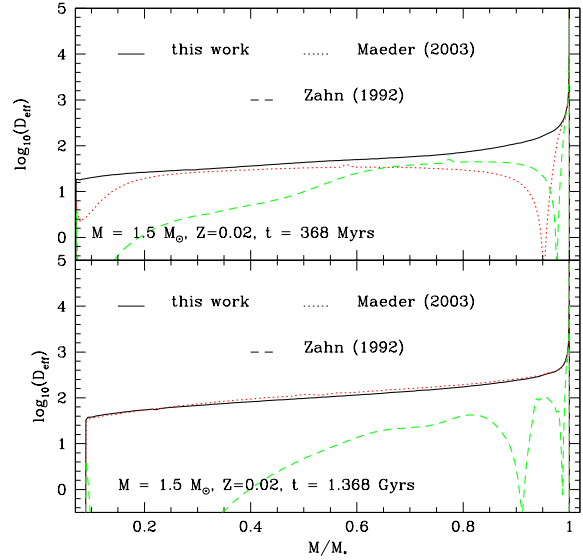}
      \caption{Same comparison for the effective diffusivity $D_{\rm eff}$ in the vertical direction, which results from the erosion of the advective transport through the horizontal diffusivity $D_h$ displayed in fig. 1. Here  $D_{\rm eff}$ has been evaluated using the old prescription for $D_h$ (Zahn 1992), that of Maeder (2003) and that of the present paper (eq. \ref{new-nuh}). 
              }
         \label{fig.2}
   \end{figure}
   
    \begin{figure}
   \centering
   \includegraphics[width=8cm]{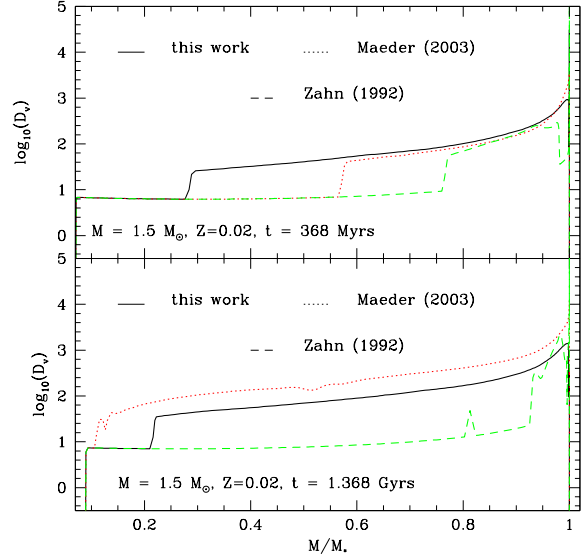}
      \caption{Same as in figs. 1 and 2 for the vertical component $D_v$ generated by the vertical shear. $D_v$ depends on $D_h$ when there is a vertical composition gradient, according to eq. \ref{eq:dv-mu}; it  has been evaluated here using the old prescription for $D_h$ (Zahn 1992), that of Maeder (2003) and that of the present paper (eq. \ref{new-nuh}). 
              }
         \label{fig.3}
   \end{figure}

\section{Comparing different prescriptions for the horizontal turbulent diffusivity}

To compare the different diffusivity prescriptions presented above, it is
necessary to implement them in a stellar structure code, since they depend
on quantities, such as the molecular weight gradient and the differential
rotation, whose evolution in time is governed by these diffusivities. For
this purpose we ran an evolutionary sequence of a 1.5~M$_\odot$ star, of
Population~I composition (X=0.705, Y=0.275, Z=0.020), using the code STAREVOL. 

The transport of chemicals and angular momentum was treated according to  
Zahn (1992) and Maeder \& Zahn (1998). The fourth-order 
  advection/diffusion equation governing the evolution of angular momentum
  in the radial direction is split into a system of four plus one first-order
  equations of order one and it is solved through a Newton-Raphson
  method. Convective zones are supposed to rotate uniformly.   We refer to Palacios et al. (2003) for more details concerning the physics used to compute of these models.

The initial equatorial velocity was taken 110 km s$^{-1}$, and the star was
submitted to magnetic spindown following Kawaler (1988) so that its
equatorial velocity at the age of the Hyades ranges between 90 and 
100 km s$^{-1}$, as expected from $v \sin i$ measurements in this
cluster (Gaig\'e 1993). The results are displayed in figs. 1-3, for two
ages: 0.368 and 1.368 Gyr.
In fig. 1 we compare the different prescriptions for the horizontal
diffusivity $D_h$, which is generated by the latitudinal shear. Note that
the more recent prescriptions, that of Maeder (eq. \ref{eq:mae-03}) and
that of the present paper (eq. \ref{new-nuh}), yield values which exceed
those of the old prescription (eq. \ref{eq:nuh-92}) by 3 to 4 orders of
magnitude.

This  enhancement of $D_h$ has two effects. First, it maintains the differential rotation in latitude at very low level; this is required by the assumption of  `shellular rotation' made in calculating the meridional circulation (Zahn 1992), namely that  to first approximation the angular velocity may be considered as a function of $r$ only. The second effect is to reduce the latitudinal fluctuations of chemical composition, which otherwise would tend to suppress the meridional circulation, as was pointed out already by Mestel (1953). The consequence is an increase of the circulation speed $U_2$, which compensates that of $D_h$, and this explains why the effective diffusivity $D_{\rm eff}$ is less affected as one would expect by the increase of $D_h$, as can be seen in fig. 2.

There are also qualitative changes in the profiles. With the $D_h$-prescription introduced here, the radial component $U_2$ of the meridional circulation keeps everywhere the same (negative) sign, and therefore $D_{\rm eff}$ never vanishes, as it does with the Maeder (2003) prescription, for which the circulation splits into three superposed cells.

Finally, this larger value of $D_h$ has the effect of increasing the turbulent diffusivity $D_v$ in the regions where it is reduced by a molecular weight gradient, according to  (\ref{eq:dv-mu}). This is illustrated in fig. 3, and one can check that $D_v$ is everywhere four orders of magnitude smaller than $D_h$, which is consistent with the anisotropy assumption made in deriving $D_{\rm eff}$.

\section{Conclusions}

We have presented here a new prescription, the  `$\beta$-viscosity',  for the horizontal component $D_h$ of the turbulent diffusivity due the differential rotation in latitude, which is drawn from Couette-Taylor experiments. According to this prescription (eq. \ref{new-nuh}),  $D_h$  exceeds by 3 or 4 orders of magnitude that which was initially suggested (eq. \ref{eq:nuh-92}) and which has been used until now. Remarkably enough, the prescription established by Maeder (2003)  yields a value which is very similar, although it is based on quite different arguments. We have analyzed the effect of this dramatic change in $D_h$ on the structure of a 1.5~M$_\odot$ star: it increases the vertical transport, and therefore enhances the mixing, particularly as the star evolves and a molecular weight gradient is building up.

The main incentive of trying to improve this prescription came from the evolutionary calculations performed by Maeder and Meynet, who found that the turbulent transport, as it was modeled initially, was insufficient to account for the observed properties of massive stars. 

This new prescription may provide a better agreement with the observations, but we can hardly consider it as the final answer, given the uncertainties involved in applying to stellar interiors a result obtained in the laboratory, in cylindrical geometry and for extreme differential rotation. However we believe that this is the best we can offer today, until numerical simulations allow us to model  that type of turbulence in a more realistic way.

\begin{acknowledgements}
The authors wish to thank A. Maeder and S. Talon for their helpful remarks. This work was partly supported by the Centre National de la Recherche Scientifique (Programme National de Physique Stellaire).  
\end{acknowledgements}

\end{document}